\documentclass[a4paper,11pt]{article}
\pdfoutput=1 

\usepackage{jheppub} 

\usepackage[T1]{fontenc} 

\usepackage{graphicx}
\usepackage{amsmath,amssymb}
\usepackage{hyperref}

\newcommand{\eepm}{\ensuremath{\mathrm{e^+e^-}}}%
\newcommand{\QQg}{\ensuremath{{Q\overline{Q}(g)}}}%
\newcommand{\QQgreal}{\ensuremath{{Q\overline{Q}g}}}%
\newcommand{\eeQQg}{\ensuremath{{\eepm\to Q\overline{Q}(g)}}}%
\newcommand{\bbg}{\ensuremath{{b\overline{b}(g)}}}%
\newcommand{\QQghard}{\ensuremath{{Q\overline{Q}g}}}%
\newcommand{\AFB}{\ensuremath{{A_{FB}}}}%
\newcommand{\AFBzeroQ}{\ensuremath{{A_{FB}^{0,Q}}}}%
\newcommand{\AFBzerob}{\ensuremath{{A_{FB}^{0,b}}}}%
\newcommand{\AFBb}{\ensuremath{{A_{FB}^{b}}}}%
\newcommand{\AFBzero}{\ensuremath{{A_{FB}^0}}}%
\newcommand{\AFBprime}{\ensuremath{{A^\prime_{FB}}}}%
\newcommand{\DeltaTwoAFB}{\ensuremath{{\Delta A^{(2)}_{FB}}}}%
\newcommand{\DeltaTwobAFB}{\ensuremath{{\Delta A^{(2,b)}_{FB}}}}%
\newcommand{\DeltaTwocAFB}{\ensuremath{{\Delta A^{(2,c)}_{FB}}}}%
\newcommand{\xbar}{\ensuremath{{\overline{x}}}}%
\newcommand{\xmin}{\ensuremath{{x_{min}}}}%
\newcommand{\xmax}{\ensuremath{{x_{max}}}}%
\newcommand{\xbmin}{\ensuremath{{\overline{x}_{min}}}}%
\newcommand{\xbmax}{\ensuremath{{\overline{x}_{max}}}}%
\newcommand{\Q}{\ensuremath{{Q}}}%
\newcommand{\Qbar}{\ensuremath{{\overline{Q}}}}%
\newcommand{\sigmaTh}{\ensuremath{{\sigma_\theta}}}%
\newcommand{\as}{\ensuremath{{\alpha_s}}}%
\newcommand{\rts}{\ensuremath{{\sqrt{s}}}}%
\newcommand{\bbbar}{\ensuremath{\mathrm{b\overline{b}}}}%
\newcommand{\MZ}{\ensuremath{m_\mathrm{Z}}}%
\newcommand{\longPar}{\ensuremath{f_L}}%
\newcommand{\cosb}{\ensuremath{\cos\theta_b}}%
\newcommand{\cosbSq}{\ensuremath{\cos^2\theta_b}}%
\newcommand{\sinbSq}{\ensuremath{\sin^2\theta_b}}%

\title{Revisiting QCD corrections to the forward-backward charge asymmetry of heavy quarks in electron-positron collisions at the Z pole: really a problem?}

\author{Juan Alcaraz Maestre}
\affiliation{CIEMAT, Avda. Complutense, 40, 28040-Madrid, SPAIN}

\abstract{We review in some detail the QCD corrections to the measurement of the forward-backward charge asymmetry of heavy quarks in the $\mathrm{e^+e^-\rightarrow Q\overline{Q}(g)}$ process at the Z pole. We show that the size of these corrections can be reduced by an order of magnitude by using simple cuts on jet acollinearity. Such a reduction is expected to lead to systematic uncertainties at the $\Delta \mathrm{A_{FB}^{0,Q}} \approx 10^{-4}$ level, opening up the path to high precision electroweak measurements with heavy flavors at future high luminosity $\eepm$ colliders like the FCC-ee.}

\begin{document} 

\maketitle

\section{Introduction}

The forward-backward asymmetry of b quarks at the Z pole, $\AFBzerob$,
is the electroweak observable that currently presents the largest
deviation with respect to standard model expectations~\cite{lephf-05}
(almost a $3\sigma$ pull). Many new-physics models have been proposed
to explain this discrepancy, all of them requiring a drastic
enhancement of the right-handed coupling of the b quark to the Z
boson~\cite{Djouadi:2006rk}. A significantly improved measurement at
the FCC-ee could thus become a clean signal of physics beyond the
standard model, provided that the deviation in the central value stays.
While the world average measurement is still dominated by statistical
uncertainties ($\Delta\AFBzerob = 0.0016$), it is also affected by
non-negligible systematic uncertainties
($\Delta\AFBzerob(\textrm{syst.})=0.0007$).

Legacy studies of the forward-backward asymmetry of heavy quarks
(bottom, charm) in $\eepm$ collisions at the Z
pole~\cite{lephf-05,ep-98} revealed that QCD corrections were the
largest source of correlated systematic uncertainty in the combined LEP
measurement. Even if some recent studies have quantified that those
uncertainties are substantially reduced when the latest parton shower
tunes are used~\cite{Enterria}, they are still expected to constitute a
dominant and irreducible source of uncertainty for high-statistics
measurements at future $\eepm$ colliders.  There have been some
attempts to reduce the size of these uncertainties, which have a
theoretical origin, through changes in the experimental analysis. For
instance, in measurements employing lepton tagging, the size of the QCD
corrections can be reduced by a factor of two or so by preferentially
selecting events with large lepton momentum~\cite{ep-98}.

The QCD corrections to the forward-backward asymmetry in $\eeQQg$ were
studied by many authors at the time of LEP/SLC
measurements~\cite{Jersak:1981sp,Djouadi:1989uk,Arbuzov:1991pr,
Djouadi:1994wt,Altarelli:1992fs,Stav:1994se}, and revisited later, with
a strong focus on the size of higher order contributions, in
References~\cite{Ravindran:1998jw,Catani:1999nf,Weinzierl:2006yt,Bernreuther:2016ccf}.
Throughout the Note, we assume that the direction of heavy quark jet
$\Q$ is the reference for the asymmetry measurement, and not the thrust
direction (which was the usual choice in LEP measurements).

In this Note we revisit this issue and explore possible approaches to
perform a measurement at future colliders with systematic uncertainties
at the $\Delta\AFB\approx 10^{-4}$ level. We first have a closer look
to the existing calculations of the QCD corrections, showing that one
can approximate them with sufficient precision using rather simple
expressions. Some relevant features, which passed almost unnoticed
until now, lead us to suggest an experimental approach using tighter
cuts in jet acollinearity than usual. This new strategy should bring a
significant reduction in the size of QCD corrections. We test the new
approach using fast simulations of the IDEA detector at the future
FCCee collider~\cite{FCC-ee}. Other potential sources of systematic
uncertainty, both of theoretical and experimental origin, are also
addressed in the study.

\section{QCD corrections at order $\mathcal{O}(\as)$: massless quark limit}

The explicit expressions for the QCD corrections at first order in
$\as$ can be found for instance in
References~\cite{Jersak:1981sp,Djouadi:1994wt}. An important feature,
not sufficiently emphasized at the time in theoretical papers, is that
virtual QCD corrections to $\AFB$ vanish in the massless quark limit.
The absence of soft-collinear divergences in the calculation implies
that the total estimate of QCD corrections is reduced to the
quantification of the corrections to the $\QQghard$ process, with one
real gluon emitted.  These hard corrections stay finite even when
approaching the configuration where the gluon is collinear with one of
the quarks or, equivalently, where the two quark jets are back-to-back.  

By convention, the QCD corrections are expressed in terms of the factor
$C(m_\Q)$:

\begin{eqnarray}
\frac{\Delta\AFB}{\AFB} & = & -\frac{\as}{\pi}~C(m_\Q) \; .
\end{eqnarray}

In the massless limit it is easy to prove that the $C$ factor is
obtained by a simple integration~\cite{Jersak:1981sp,Djouadi:1994wt}:

\begin{eqnarray}
C(0) & = & \frac{4}{3}~\int_{\xmin}^{\xmax} dx \int_{\xbmin(x)}^{\xbmax(x)} d\xbar~\left(\frac{\xbar}{x}\right) \; , \label{eq:C-nomass}
\end{eqnarray}

\noindent 
where $x$ and $\xbar$ are the reduced energies of the heavy quarks in
the $\eepm\to\QQg$ process: $x\equiv 2E_\Q/\rts$, $\xbar\equiv
2E_\Qbar/\rts$. When integrated over the full available phase space,
the limits of the integral are: $\xbmin(x)=1-x,\xbmax(x)=1,\xmin=0,
\xmax=1$, leading to the known result:

\begin{eqnarray}
C(0,\text{full phase space}) & = & \frac{4}{3}~\int_{0}^{1} dx \int_{1-x}^{1} d\xbar~\left(\frac{\xbar}{x}\right) = 1\; .
\end{eqnarray}

In this context, it is possible to apply cuts on observables that
depend on $x$ and $\xbar$. One example is the acollinearity $\xi$
between quark and antiquark directions~\cite{Djouadi:1994wt}:

\begin{eqnarray}
\cos\xi(x,\xbar) & = & 1 - 2~\frac{(1-x)~(1-\xbar)}{x~\xbar} \\
& \Downarrow & \nonumber \\
{\displaystyle \sin^2(\xi(x,\xbar)/2)} & = & \frac{(1-x)~(1-\xbar)}{x~\xbar} \; .
\end{eqnarray}

The limits for an acollinearity cut $\xi<\xi_0$ are
$\xbmin(x)=(1-x)/(1-x+\epsilon_0 x)$, $\xbmax(x)=1$, $\xmin=0,\xmax=1$,
where we define: $\epsilon_0\equiv\sin^2(\xi_0/2)$. With this we obtain
the more general expression:

\begin{eqnarray}
\epsilon_0 & \equiv & \sin^2(\xi_0/2) \\
& \Downarrow & \nonumber \\
C(0,\xi<\xi_0) & = & \int_{0}^{1} dx \int_{\frac{1-x}{1-x+\epsilon_0 x}}^{1} d\xbar~\left(\frac{4\xbar}{3x}\right) = {\displaystyle \frac{-2\epsilon_0\left[(2-\epsilon_0) (\log\epsilon_0)+1-\epsilon_0 \right]}{3(1-\epsilon_0)^2}} \; . \label{eq:C-nomass-acol}
\end{eqnarray}

From Equation~\ref{eq:C-nomass-acol} we derive the results listed in
Table~\ref{tab:C-nomass}. They show that the correction factor can be
significantly reduced with the use of acollinearity cuts. For instance,
for $\xi_0=0.3$, which is still larger than the typical angular
resolution of experimental jets, C(0) is reduced by one order of
magnitude.

\begin{table}[htb]
\begin{center}

\begin{tabular}{|c|c|} \hline
$\xi_0$ cut & $C(0,\xi<\xi_0)$ \\
\hline\hline
No cut & $1.000$ \\
$1.50$ & $0.693$ \\
$1.00$ & $0.473$ \\
$0.50$ & $0.207$ \\
$0.30$ & $0.102$ \\
$0.20$ & $0.055$ \\
$0.10$ & $0.018$ \\
$0.05$ & $0.006$ \\
\hline
\end{tabular}

\caption{QCD correction factor $C$ in the massless quark limit
($m_\Q=0$), for different acollinearity cuts $\xi<\xi_0$ between the
two quarks in the $\eepm\to\QQg$ process.}

\label{tab:C-nomass}

\end{center}
\end{table}

\section{QCD corrections at order $\mathcal{O}(\as)$: the massive case ($m_\Q\ne 0$) \label{sec:C-massive}}

In the massive case, even at order $\as$, virtual QCD corrections to
the asymmetry are not zero, and soft, virtual and hard corrections must
be added together in order to obtain finite results. The first order
corrections can be expanded in powers of the small parameter $\mu\equiv
2~m_\Q/\rts$~\cite{Stav:1994se} ($\mu\approx 0.1$ at the Z pole for
b-quarks). Fortunately, the explicitly divergent terms lead to
corrections of order $\mu^2\approx 0.01$, and the leading corrections
can be calculated to a good approximation by just updating to the
massive case the limits of integration in Equation~\ref{eq:C-nomass}.
This key detail at order $\as$ was already noticed by the authors of
Reference~\cite{Altarelli:1992fs}. Accordingly, we can obtain an
approximation to $C(m_\Q)$ at the percent level from rather simple
numerical integrations. This precision is sufficient for our purposes.
For instance, let us note that the ${\cal O}(\as^2$) corrections to the
asymmetry~\cite{Altarelli:1992fs,Ravindran:1998jw,Catani:1999nf,Weinzierl:2006yt,Bernreuther:2016ccf}
are at least one order of magnitude larger than the uncertainties due
to the approximation employed here.

In the massive case the acollinearity angle between the two quarks in
the $\QQgreal$ final state has the following dependence on $x,\xbar$
and $\mu$~\cite{Djouadi:1994wt}:

\begin{eqnarray}
\cos\xi(x,\xbar,\mu) & = & \frac{x\xbar - \mu^2 - 2~(1-x)~(1-\xbar)}{\sqrt{x^2-\mu^2}~\sqrt{\xbar^2-\mu^2}}.
\end{eqnarray}

Taking the integrand for the hard QCD corrections without
additional $\mu^2$ terms we get:

\begin{eqnarray}
C(\mu) & \approx & \int_{\xmin}^{\xmax} dx \int_{\xbmin(x)}^{\xbmax(x)} d\xbar~\left[\frac{2\xbar^2(1-\cos\xi(x,\xbar,\mu))}{3(1-x)(1-\xbar)}\right] \; , \label{eq:C-mass}
\end{eqnarray}

\noindent which by construction agrees with Equation~\ref{eq:C-nomass}
in the massless limit. For the full phase space, the $\xbmin$ limit is
reached in the configuration where the antibottom quark is emitted
along the direction of the bottom quark, while the $\xbmax$ limit is
reached when they are back-to-back. In addition, the minimum reduced
energy of a quark is $\mu$. This gives:

\begin{eqnarray}
\xbmin(x) & = & 1 - \frac{x+\sqrt{x^2-\mu^2}}{2} + \frac{\mu^2}{2-x-\sqrt{x^2-\mu^2}} \label{mulimits-first} \\
\xbmax(x) & = & 1 - \frac{x-\sqrt{x^2-\mu^2}}{2} + \frac{\mu^2}{2-x+\sqrt{x^2-\mu^2}} \\
\xmin & = & \mu \\
\xmax & = & 1 \;. \label{mulimits-last}
\end{eqnarray}

The numerical integration of Equation~\ref{eq:C-mass} over the full
phase space leads to the values presented in
Table~\ref{tab:C-mass-noacol}. They are in excellent agreement with
those quoted in Reference~\cite{Djouadi:1994wt} for several input
masses of the charm ($0.7$, $1.5$~GeV) and bottom ($3.0$, $4.5$~GeV)
quarks, thus confirming the expected percent accuracy of the
approximation. As commented before, the main quantitative difference in
the calculation with respect to the massless case is the modification
of the limits of integration. Therefore, and to the $\mu^2$ level of
proposed precision, we could have used instead the integrand of
Equation~\ref{eq:C-nomass}. The results of this alternative
approximation, although not shown, also agree as well at the percent
level with the exact results.  Let us finally note that our calculation
does not have any explicit dependence on the values of the vector and
axial couplings of the Z boson to fermions. This is so because those
dependencies only appear at order $\mu^2$.

\begin{table}[htb]
\begin{center}

\begin{tabular}{|c|c|c|} \hline
$m_\Q$~[GeV] & $C$ (our approximation) & $C$ (Ref.\cite{Djouadi:1994wt}) \\
\hline\hline
$0.0$ & $1.00$ & $1.00$ \\
$0.7$ & $0.96$ & $0.96$ \\
$1.5$ & $0.92$ & $0.93$ \\
$3.0$ & $0.86$ & $0.86$ \\
$4.5$ & $0.80$ & $0.80$ \\
\hline
\end{tabular}

\caption{QCD correction factor $C$ for the massive case and full phase
space using the ${\cal O}(\as)$ approximation described in the text.
The obtained values are found to be in excellent agreement with those
obtained via numerical integration of the exact expressions in
Reference~\cite{Djouadi:1994wt}.}

\label{tab:C-mass-noacol}

\end{center}
\end{table}

It is relatively easy to determine the corresponding corrections in the
presence of an acollinearity cut by imposing a cutoff
$\cos\xi(x,\xbar,\mu)>\cos\xi_0$ to the integrand: 

\begin{eqnarray}
C(\mu,\xi_0) \approx \int_{\xmin}^{\xmax} dx \int_{\xbmin(x)}^{\xbmax(x)} d\xbar~\left[\frac{2\xbar^2(1-\cos\xi(x,\xbar,\mu))}{3(1-x)(1-\xbar)}\right] \Theta\left(\frac{\cos\xi(x,\xbar,\mu)}{\cos\xi_0}-1\right).
\end{eqnarray}

The results of the calculations for the heavy-quark masses
$m_\Q=0.7,1.5,3.0,4.5$~GeV are shown in Table~\ref{tab:C-mass-acol},
compared with those obtained in the massless case. The main conclusion
is that the same trend is maintained in the massive case. A simple cut
$\xi<0.3$ is again sufficient to reduce the size of QCD corrections by
one order of magnitude.

\begin{table}[htb]
\begin{center}

\begin{tabular}{|c|c|c|c|} \hline
$\xi_0$ cut & $C_{\xi<\xi_0}$, $m_c=0$ & $C_{\xi<\xi_0}$, $m_c=0.7$~GeV & $C_{\xi<\xi_0}$, $m_c=1.5$~GeV \\
\hline\hline
No cut & $1.00$ & $0.96$ & $0.92$ \\
$1.50$ & $0.69$ & $0.67$ & $0.65$ \\
$1.00$ & $0.47$ & $0.46$ & $0.44$ \\
$0.50$ & $0.21$ & $0.20$ & $0.19$ \\
$0.30$ & $0.10$ & $0.10$ & $0.09$ \\
$0.20$ & $0.06$ & $0.05$ & $0.04$ \\
$0.10$ & $0.02$ & $0.01$ & $0.01$ \\
\hline
\end{tabular}

\vspace{0.02\textheight}

\begin{tabular}{|c|c|c|c|} \hline
$\xi_0$ cut & $C_{\xi<\xi_0}$, $m_b=0$ & $C_{\xi<\xi_0}$, $m_b=3.0$~GeV & $C_{\xi<\xi_0}$, $m_b=4.5$~GeV \\
\hline\hline
No cut & $1.00$ & $0.86$ & $0.80$ \\
$1.50$ & $0.69$ & $0.61$ & $0.57$ \\
$1.00$ & $0.47$ & $0.42$ & $0.39$ \\
$0.50$ & $0.21$ & $0.18$ & $0.17$ \\
$0.30$ & $0.10$ & $0.09$ & $0.08$ \\
$0.20$ & $0.06$ & $0.04$ & $0.04$ \\
$0.10$ & $0.02$ & $0.01$ & $0.01$ \\
\hline
\end{tabular}

\caption{QCD correction factor $C$ for different heavy quark masses
and quark acollinearity ($\xi_0$) cuts in the $\eepm\to\QQg$ process.}

\label{tab:C-mass-acol}

\end{center}
\end{table}

\section{Other distortions in the angular distribution due to QCD effects \label{sec:fL}}

QCD radiation effects also modify the forward-backward symmetric part
of the angular distribution. In the absence of hard non-collinear
radiation effects and neglecting the contributions from box diagrams
and initial-final state interferences at the Z pole, the differential
polar angular distribution is given by the general expression:

\begin{eqnarray}
      \frac{d\rho}{d\cosb} & = & \frac{1}{1+\longPar}~\left[\frac{3}{8}~(1+\cosbSq)+\frac{3}{4}~\longPar\sinbSq \right]+\AFB~\cosb \; . \label{eq:exact-angular-distribution}
\end{eqnarray}

At Born level, the ``longitudinal'' fraction $\longPar$ is always
smaller than $\mu^2/2\lesssim 0.005$, too small to have a significant
impact on an experimental measurement with $\Delta\AFB/\AFB\gtrsim
0.1\%$. However, a non-negligible longitudinal component develops after
the inclusion of QCD radiation effects. At order $\mathcal{O}(\as)$,
and in the $\mu\to 0$ limit, the resulting correction has a simple
analytical expression, as derived from the calculations of
Reference~\cite{Jersak:1981sp}:

\begin{eqnarray}
\longPar(\mu=0,\xi_0) & = & \int_{0}^{1} dx \int_{\frac{1-x}{1-x+\epsilon_0 x}}^{1} d\xbar~\frac{4\as(x+\xbar-1)}{3\pi x^2} = -\frac{2\as\epsilon_0}{3\pi}~\left[\frac{2\epsilon_0\log(\epsilon_0)}{(1-\epsilon_0)}+1\right],
\end{eqnarray}

\noindent where again $\epsilon_0=\sin^2(\xi_0/2)$. The results for
different values of the acollinearity cut $\xi_0$ are shown in the
second column of Table~\ref{tab:epsilonL-theo}.  In the absence of
acollinearity cuts ($\epsilon_0=1$),
$\longPar=\frac{\as}{\pi}r_L=\frac{\as}{\pi}\frac{2}{3}\approx 0.025$,
in exact agreement with the result: $r_L=r^\prime_L=\frac{2}{3}$ from
Reference~\cite{Jersak:1981sp} in the $\mu\to 0$
limit~\footnote{Differences between the parameters $r_L$ and
$r^\prime_L$ appear only at order $\mathcal{O}(\mu^2)\lesssim 1\%$, and
are therefore negligible in practice for charm and bottom quarks.}.  Note
that $\longPar$ vanishes in the zero-acollinearity limit. For values
$\longPar\lesssim 0.005$ the bias in a measurement that ignores the
longitudinal component is: $\delta\AFB/\AFB\lesssim 0.1\%$, as
estimated using toy samples generated according to the true
distribution and fitted to an angular distribution with $\longPar=0$.
The table indicates that such an approximation is appropriate for
acollinearity cuts $\xi_0\lesssim 0.3$.  

\begin{table}[htb]
\begin{center}

\begin{tabular}{|c|c|c|c|} \hline
$\xi_0$ cut & $\longPar$, $m_\Q=0$ & $\longPar$, $m_\Q=1.5$~GeV & $\longPar$, $m_\Q=4.5$~GeV \\
\hline\hline
No cut & $0.025$ & $0.023$ & $0.020$ \\
$1.50$ & $0.022$ & $0.020$ & $0.018$ \\
$1.00$ & $0.016$ & $0.015$ & $0.014$ \\
$0.50$ & $0.008$ & $0.007$ & $0.006$ \\
$0.30$ & $0.004$ & $0.003$ & $0.003$ \\
$0.20$ & $0.002$ & $0.002$ & $0.002$ \\
$0.10$ & $0.001$ & $0.000$ & $0.000$ \\
\hline
\end{tabular}

\caption{Fraction of the longitudinal component of the polar angle differential distribution for different values of the heavy quark mass
and the acollinearity cut ($\xi_0$) in the $\eepm\to\QQg$ process.}

\label{tab:epsilonL-theo}

\end{center}
\end{table}

Expectations for the massive case at the percent (relative) level can
be obtained by ignoring additional $\mathcal{O}(\mu^2)$ terms in the
integrand, following the same strategy employed in Section~\ref{sec:C-massive}. The explicit expression for $\longPar$ is:

\begin{eqnarray}
\longPar(\mu,\xi_0) \approx \int_{\xmin}^{\xmax} dx \int_{\xbmin(x)}^{\xbmax(x)} d\xbar~\frac{4\as\sqrt{\xbar^2-\mu^2} \sin^2\xi(x,\xbar,\mu)}{3\pi (1-x)(1-\xbar)} \Theta\left(\frac{\cos\xi(x,\xbar,\mu)}{\cos\xi_0}-1\right)
\end{eqnarray}

\noindent with the integration limits provided by
Equations~\ref{mulimits-first}-\ref{mulimits-last}. The results of a
numerical integration for the values of the heavy quark masses
$m_\Q=1.5, 4.5$~GeV are given in Table~\ref{tab:epsilonL-theo},
confirming the tendency observed in the massless case. Again, there is
good agreement with the mass dependence of the $r_L,r^\prime_L$
parameters shown graphically in Reference~\cite{Jersak:1981sp}.
Similarly to the $C$ parameter case discussed in
Section~\ref{sec:C-massive}, the estimates for $\longPar$ do not depend
on the flavor of the quark, because the correction has no dependence on
the Z boson couplings at this level of approximation.

\section{QCD corrections at order $\mathcal{O}(\as^2)$ and mass uncertainties \label{sec:as2}}

The first calculation of the QCD corrections to order $\as^2$ was
performed in Reference~\cite{Altarelli:1992fs} for the massless case.
The inclusion of second order corrections led to additional negative
shifts of $\DeltaTwobAFB/\AFB \approx -0.003$ for bottom quarks and
$\DeltaTwocAFB/\AFB \approx -0.006$ for charm quarks. A subsequent
calculation~\cite{Ravindran:1998jw}, quantitatively confirmed by
Reference~\cite{Catani:1999nf}, obtained significantly larger
contributions, $\DeltaTwobAFB/\AFB \approx -0.012$, $\DeltaTwocAFB/\AFB
\approx -0.016$. The most recent calculation of
Reference~\cite{Bernreuther:2016ccf}, considering for the first time
the fully massive b-quark case, estimates a shift of
$\DeltaTwobAFB/\AFB \approx -0.005$ for $m_b=4.89$~GeV and is
consistent with the results of
References~\cite{Ravindran:1998jw,Catani:1999nf} for the massless case.
The dominant source of uncertainty in the calculation of the full QCD
correction factor at order $\mathcal{O}(\as^2)$ is the arbitrariness in
the choice of the renormalization scale. Its effect on the asymmetry is
not negligible: $\Delta\AFB/\AFB \approx \pm
0.003$~\cite{Bernreuther:2016ccf}.

In the previous numbers we explicitly neglect the so-called ``singlet''
contributions, which concern 4-quark final states where at least two of
the quarks are heavy and can be interpreted as originating from
gluon-splitting. These contributions are largely dependent on
experimental cuts and are usually considered as a separate
(non-dominant) source of uncertainty~\cite{lephf-05}. The inclusion of
singlet contributions in the theoretical estimates without cuts would
imply second order corrections as large as $\approx 40\%$ times the
size of the first order corrections~\cite{Bernreuther:2016ccf}.

Nevertheless, and according to the discussion of the previous sections,
it is expected that the application of tighter cuts on acollinearity
will bring these corrections to the $\DeltaTwoAFB/\AFB \approx -0.001$
level. Last but not least, a better, infrared-safe definition of
flavored quark jets~\cite{Banfi:2006hf} could be used in the
measurement. This alternative definition is known to further reduce the
size of QCD corrections, as shown in Reference~\cite{Weinzierl:2006yt}.

In theoretical estimates at order $\mathcal{O}(\as)$, the values of
heavy-quark masses are typically varied from their value in the
$\overline{\text{MS}}$ scheme at the $\MZ$ scale to their pole value.
These variations translate into effects as large as $10\%$ on $C$, i.e.
$\Delta\AFB/\AFB \approx \pm 0.004\times C$. Therefore, a reduction of
the $C$ value by an order of magnitude using acollinearity cuts should
bring the uncertainty to the $\Delta\AFB/\AFB \approx 0.0004$ level.
Mass uncertainties are much reduced in calculations at order
$\mathcal{O}(\as^2)$ with massive quarks~\cite{Bernreuther:2016ccf},
which properly take into account the evolution of the heavy quark mass
with the renormalization scale.

Estimates of the longitudinal fraction $f_L$ are also available at
order $\mathcal{O}(\as^2)$ for the massless case.  From the
calculations of Reference~\cite{Ravindran:1998jw} we deduce a
longitudinal fraction for the bottom quark case of: $\longPar =
0.034\pm 0.001$, where the uncertainty covers running quark mass
effects. There is a $\approx 0.01$ difference between this result and
the $\mathcal{O}(\as)$ result, but it only translates into an
uncertainty of $\Delta\AFB/\AFB \approx \pm 0.002$. Differences are
expected to be significantly smaller in the presence of acollinearity
cuts.

In summary, and from a purely theoretical point of view, the
application of an acollinearity cut of the type $\xi\lesssim 0.3$
should provide a measurement of the bare forward-backward asymmetry
with relative biases due to QCD corrections below $0.01$ and relative
uncertainties at the permille level.

\section{QCD corrections in parton-shower simulated events}

Despite the availability of precise $\mathcal{O}(\as^2)$ theoretical
estimates at parton level, an accurate estimate of the size of QCD
corrections requires a dedicated study using parton-shower generators
and explicit experimental cuts~\cite{ep-98}.  In order to explore the
possibility of an experimental measurement of the forward-backward
asymmetry using acollinearity cuts, we fast-simulate 10 million
$\eepm\to\bbbar$ events at the Z pole, $\rts = 91.2$~GeV, using the
FCC-ee implementation of the IDEA detector~\cite{FCC-ee} in
DELPHES\cite{DELPHES}. The generator interface with the DELPHES
simulator is PYTHIA~\cite{PYTHIA6,PYTHIA8}.  Regarding hadronization
and timelike-showering aspects we use as central value the Monash 2013
tune~\cite{MONASH-2013}, although all the available alternative choices
will be used later to quantify systematic uncertainties due to this
kind of effects.

\subsection{Effect of acollinearity cuts at generator level}

At the generator level in this sample, the forward-backward asymmetry
before parton showering, fragmentation and hadronization is found to
be: $\AFBzero = 0.1036 \pm 0.0003$.  This number already includes
initial-state QED corrections. After the inclusion of QCD effects, the
asymmetry, measured in the full phase space (no cuts) becomes:
$\AFB(\xi<\pi) = 0.0996 \pm 0.0003$.  The generator-level asymmetry for
different acollinearity cuts between bottom and antibottom quarks is
shown in Table~\ref{tab:afb-gen}. There is a nice asymptotic tendency
to recover (within uncertainties) the asymmetry value without QCD
corrections at tight values of the cut.  The statistical uncertainty on
the observed shifts is small because basically the same events are used
in the generator-level calculations of $\AFB$ and $\AFBzero$ for each
acollinearity cut. Possible systematic uncertainties due to
parton-shower uncertainties are discussed in Section~\ref{sec:tunes}.
The expected shifts are calculated for a value of $m_b=4.5$~GeV and
assume a $\approx 10\%$ relative uncertainty from mass quark
uncertainties (see Section~\ref{sec:as2}).  Let us point out again that
the QCD corrections at the parton-shower level are not expected to
exactly match those theoretical expectations~\cite{ep-98}.

\begin{table}[htb]
\begin{center}

\begin{tabular}{|c|c|c|c|} \hline
$\xi_0$ cut & $\AFB (\text{Pythia}, \xi<\xi_0)$ & $\AFB-\AFBzero (\text{Pythia})$ & Expected shift at $\mathcal{O}(\as)$ \\
\hline\hline
No cut & $0.0996 \pm 0.0002(\mathrm{stat.})$ & $-0.0040$ & $-0.00312 \pm 0.00031(\mathrm{theo.})$ \\
$1.50$ & $0.1007 \pm 0.0002(\mathrm{stat.})$ & $-0.0029$ & $-0.00220 \pm 0.00022(\mathrm{theo.})$ \\
$1.00$ & $0.1016 \pm 0.0002(\mathrm{stat.})$ & $-0.0021$ & $-0.00154 \pm 0.00015(\mathrm{theo.})$ \\
$0.50$ & $0.1029 \pm 0.0002(\mathrm{stat.})$ & $-0.0010$ & $-0.00067 \pm 0.00007(\mathrm{theo.})$ \\
$0.30$ & $0.1034 \pm 0.0002(\mathrm{stat.})$ & $-0.0005$ & $-0.00032 \pm 0.00003(\mathrm{theo.})$ \\
$0.20$ & $0.1033 \pm 0.0003(\mathrm{stat.})$ & $-0.0003$ & $-0.00016 \pm 0.00002(\mathrm{theo.})$ \\
$0.10$ & $0.1038 \pm 0.0003(\mathrm{stat.})$ & $-0.0002$ & $-0.00005 \pm 0.00001(\mathrm{theo.})$ \\
\hline
\end{tabular}

\caption{Forward-backward asymmetry evaluated at the generator level in
$10^7$ $\eepm\to\bbbar(\mathrm{g})$ events at $\rts = 91.2$~GeV, for
different cuts of the acollinearity ($\xi<\xi_0$) between the two
bottom quarks. The observed and $\mathcal{O}(\as)$ expected shifts with
respect to $\AFBzero$ due to theoretical QCD corrections are also
shown. The statistical uncertainties on the observed $\AFB-\AFBzero$
differences are negligible because the same events are used in the
determination of $\AFB$ and $\AFBzero$ for each acollinearity cut.
Systematic uncertainties are not considered at this level. The
theoretical values assume a $\approx 10\%$ relative uncertainty.}

\label{tab:afb-gen}

\end{center}
\end{table}

We also compute the value of the longitudinal fraction $\longPar$ as a
function of the $\xi_0$ cut. For convenience, it is extracted from the
average value of $|\cosb|$, $<|\cosb|>$, which is connected to
$\longPar$ via the relation:

\begin{eqnarray}
      \longPar & = & \frac{9-16<|\cosb|>}{-6+16<|\cosb|>} \\
\end{eqnarray}

The results are presented in Table~\ref{tab:longPar} for the Monash
2013 tune, although no significant dependence on the choice of tune is
observed. The statistical uncertainty is derived from the variance of
$<|\cosb|>$. As expected, longitudinal effects become almost negligible
for small values of the acollinearity cut.  The value in the absence of
cuts, $\longPar = 0.029\pm 0.001~(\mathrm{stat.})$, can be compared
with the $\mathcal{O}(\as^2)$ prediction for the bottom case extracted
from Reference~\cite{Ravindran:1998jw}, $\longPar = 0.034\pm 0.001$.
Even if small, the discrepancy between between the two values can be
explained by the intrinsic differences between the treatments (massless
calculation at $\mathcal{O}(\as^2)$, missing corrections of higher
order, hard-non collinear effects, \ldots).  In any case, it is
important to point out that the level of disagreement is already of the
order of the accuracy required for an experimental measurement with
precision $\Delta\AFBb\approx 0.0001$.

\begin{table}[htb]
\begin{center}

\begin{tabular}{|c|c|} \hline
$\xi_0$ cut & $\longPar$ for $\xi<\xi_0$ \\
\hline\hline
No cut & $0.029 \pm 0.001(\text{stat.})$ \\
$1.50$ & $0.026 \pm 0.001(\text{stat.})$ \\
$1.00$ & $0.021 \pm 0.001(\text{stat.})$ \\
$0.50$ & $0.011 \pm 0.001(\text{stat.})$ \\
$0.30$ & $0.006 \pm 0.001(\text{stat.})$ \\
$0.20$ & $0.004 \pm 0.001(\text{stat.})$ \\
$0.10$ & $0.001 \pm 0.001(\text{stat.})$ \\
\hline
\end{tabular}

\caption{Estimated values of the $\longPar$ parameter in $10^7$
Pythia $\eepm\to\bbbar(\mathrm{g})$ events generated at the Z pole, for
different cuts of the acollinearity ($\xi<\xi_0$) between the two
bottom quarks. Only statistical uncertainties are given.}

\label{tab:longPar}

\end{center}
\end{table}

\subsection{Effect of acollinearity cuts at reconstructed level \label{sec:afb-rec}}

As a first realistic test, we measure the asymmetry using events with
at least two reconstructed b-tagged jets. These jets have particle-flow
candidates as constituents~\cite{FCC-ee}, and are defined using an
anti-$k_T$ $\eepm$ generalized algorithm~\cite{Cacciari:2011ma} with
distance parameter $R=0.4$.  Given the main purpose of the exercise -
testing QCD corrections in the presence of acollinearity cuts - we
simplify the analysis assuming an ideal charge assignment to the jet
associated to the b-quark. On the other hand, and in order to avoid
biases due to changes in the jet acceptance as a function of the polar
angle, the asymmetry is determined via a likelihood fit to the expected
angular distribution given by
Equation~\ref{eq:exact-angular-distribution}.  The fit is insensitive
to acceptance or efficiency corrections that are dependent on the polar
angle, as long as they are charge symmetric. It also provides by
construction the smallest statistical variance for the measurement in
the limit of large number of events. We account for the presence of a
$\sin^2\theta$ longitudinal component by fixing the estimated
generator-level $\longPar$ fractions to the values given in
Table~\ref{tab:longPar}. The final result is corrected as well for
biases due to angular resolution effects.

\begin{figure}[htbp]
\centering
\includegraphics[width=0.45\linewidth]{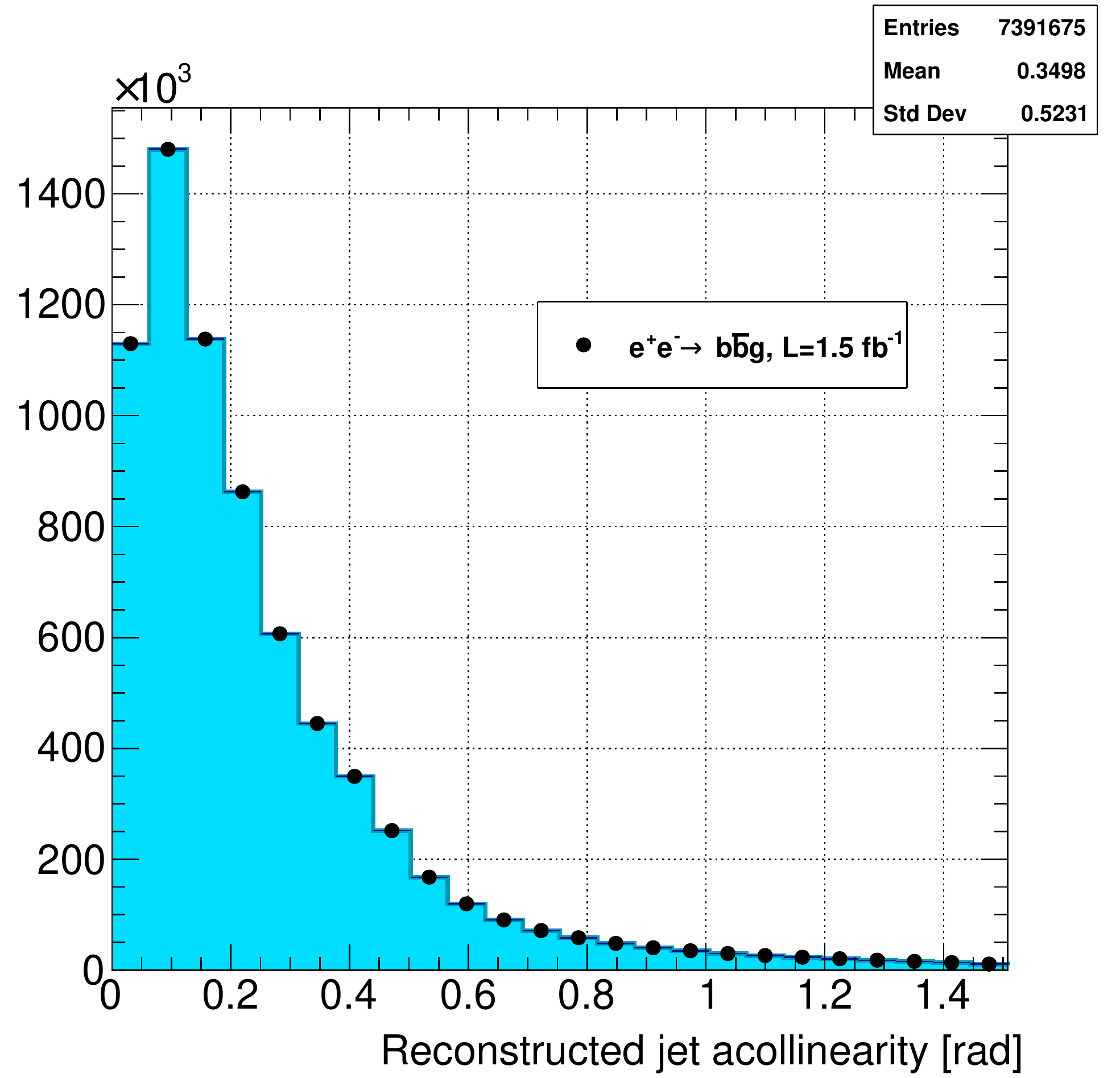}
\includegraphics[width=0.45\linewidth]{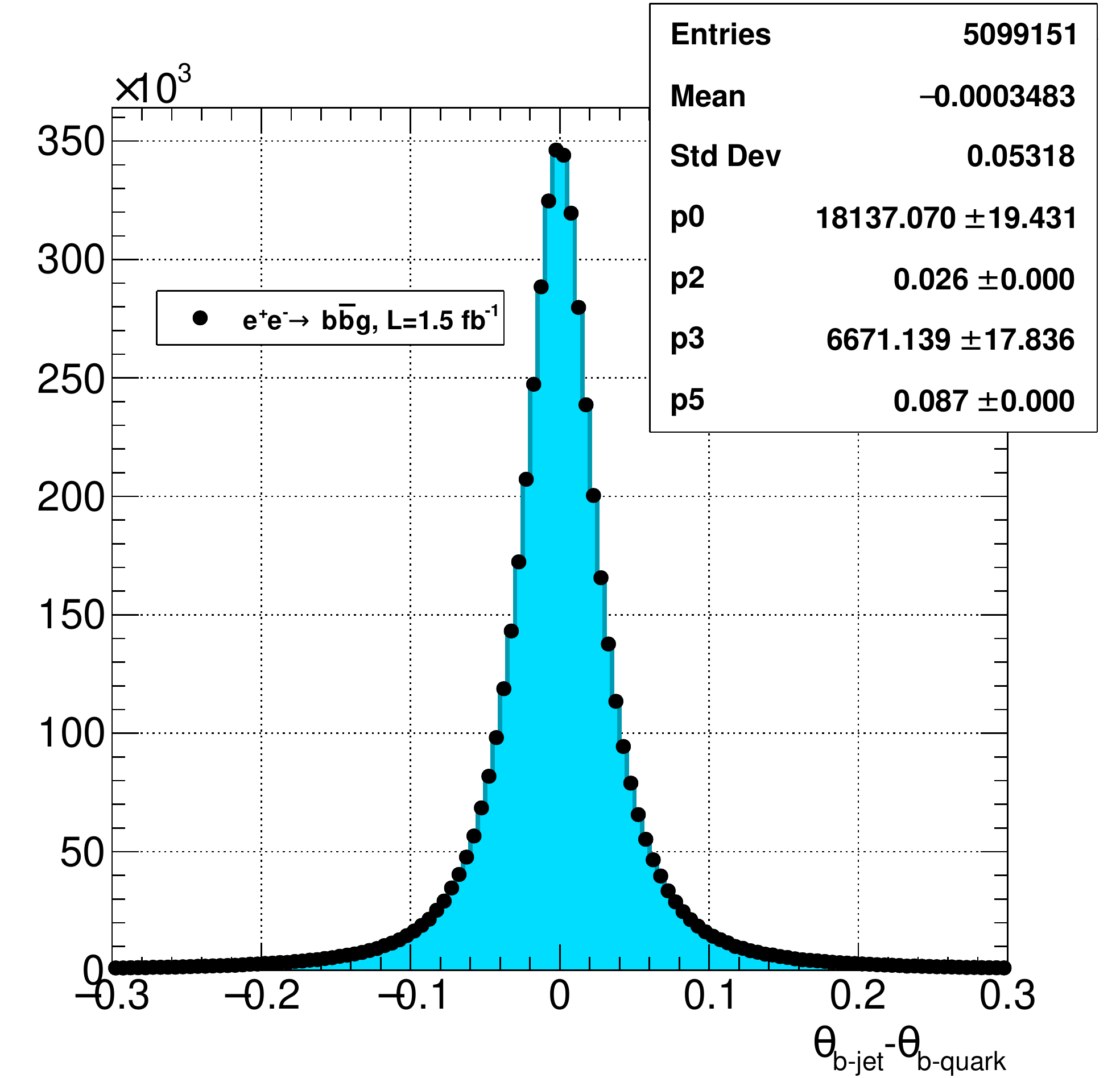}

\caption{ Left: acollinearity distribution between the two b-tagged
reconstructed jets in selected $\eepm\to\bbg$ events.  Right: angular
resolution on the original b-quark direction using reconstructed
b-tagged jets, for an acollinearity cut of $\xi<0.3$.  A fast
simulation with DELPHES of an IDEA detector at the FCC-ee is used in
the study \label{fig:angular-reso}.}

\end{figure}

For each event in the pseudo-data sample, we define the observed
acollinearity $\xi$ as $\pi$ minus the maximum angular distance between
the reconstructed  b-jet and any other b-tagged jet in the event. A
simple inspection of its distribution
(Figure~\ref{fig:angular-reso}-left) is enough to conclude that large
data samples are still available for analysis with cuts as low as
$\xi<0.2-0.3$.  Figure~\ref{fig:angular-reso}-right shows the angular
distance between the charge-tagged b-quark jet and the original bottom
quark at the generator level for a $\xi<0.3$ cut. An r.m.s. of $\approx
0.05$ radians is observed. When fitted to two Gaussian components, we
obtain a first component with width $~\approx 0.026$~rad for about
$73\%$ of the events, and a wider component of width $\approx
0.087$~rad for the remaining fraction. This angular resolution produces
minimal biases in the measurement, which are discussed in more detail
in Section~\ref{sec:bias}.

The measured values of the asymmetry for different $\xi<\xi_0$ cuts are
reported in Table~\ref{tab:afb-rec}~\footnote{The results also include
a small correction due to the limited angular resolution (see
Section~\ref{sec:bias}).}. The last column of the table contains the
results after correction of the QCD shifts calculated at generator
level (third column of Table~\ref{tab:afb-gen}). We observe that the
measured asymmetry approaches with enough precision $\AFBzero$ at
relatively low values of $\xi_0$, $\xi_0 \approx 0.2-0.3$, for which
QCD uncertainties should be almost negligible. There is a significant
deviation from expectations for $\xi_0=0.1$. This value is not far from
the estimated r.m.s. of the polar angle resolution, which may provoke
distortions in the fitted distribution that we have not explored yet.
Also, results for too low values of $\xi_0$ can be affected by
intra-jet systematics that are difficult to evaluate at this stage.

\begin{table}[htb]
\begin{center}

\begin{tabular}{|c|c|c|} \hline
$\xi_0$ cut & $\AFB (\xi<\xi_0)$ & $\AFB$ QCD-corrected \\
\hline\hline
No cut & $0.0992 \pm 0.0002(\mathrm{stat.})$ & $0.1032 \pm 0.0002(\mathrm{stat.})$\\
$1.50$ & $0.0997 \pm 0.0003(\mathrm{stat.})$ & $0.1026 \pm 0.0003(\mathrm{stat.})$\\
$1.00$ & $0.1006 \pm 0.0003(\mathrm{stat.})$ & $0.1027 \pm 0.0003(\mathrm{stat.})$\\
$0.50$ & $0.1022 \pm 0.0003(\mathrm{stat.})$ & $0.1032 \pm 0.0003(\mathrm{stat.})$\\
$0.30$ & $0.1028 \pm 0.0003(\mathrm{stat.})$ & $0.1033 \pm 0.0003(\mathrm{stat.})$\\
$0.20$ & $0.1031 \pm 0.0003(\mathrm{stat.})$ & $0.1034 \pm 0.0003(\mathrm{stat.})$\\
$0.10$ & $0.1021 \pm 0.0004(\mathrm{stat.})$ & $0.1023 \pm 0.0004(\mathrm{stat.})$\\
\hline
\end{tabular}

\caption{Forward-backward asymmetry measured in $10^7$ simulated
$\eepm\to\bbbar(\mathrm{g})$ events at $\rts = 91.2$~GeV, for different
$\xi<\xi_0$ cuts between the two selected b-tagged reconstructed jets.
The last column provides those asymmetry values corrected by QCD
effects. The expected asymmetry in the absence of QCD corrections is:
$\AFBzero = 0.1036 \pm 0.0003$. Only statistical uncertainties are
shown.}

\label{tab:afb-rec}

\end{center}
\end{table}

\subsection{Stability under hadronization and timelike-showering uncertainties \label{sec:tunes}}

The same simplified study is repeated on 6 additional samples,
generated according to the alternative shower tunes available in
PYTHIA8~\cite{PYTHIA8} for $\eepm$ collisions. Similarly to the
reference study, 10 million $\eepm\to\bbbar(g)$ events at the Z pole
were generated for each tune. For reference, the available choices are
listed in Table~\ref{tab:tunes}. 

\begin{table}[htb]
\begin{center}

\begin{tabular}{|c|l|} \hline
Pythia8 $\eepm$ tune & Description \\
\hline\hline

1 & the original PYTHIA 8 parameter set, based on some very old \\
  & flavour studies (with JETSET around 1990) and a simple tune \\
  & of $\as$ to three-jet shapes to the new $p_T$-ordered shower. \\
\hline
2 & a tune by Marc Montull to the LEP 1 particle composition, as \\
  & published in the RPP (August 2007). No related (re)tune to \\
  & event shapes has been performed, however. \\
\hline
3 & a tune to a wide selection of LEP1 data by Hendrik Hoeth \\
  & within the Rivet + Professor framework, both to hadronization \\
  & and timelike-shower parameters (June 2009). \\
\hline
4 &  a tune to LEP data by Peter Skands, by hand, both to \\
  & hadronization and timelike-shower parameters (September 2013). \\
\hline
5 & first tune to LEP data by Nadine Fischer (September 2013), \\
  & based on the default flavour-composition parameters. Input \\
  & is event shapes (ALEPH and DELPHI), identified particle \\
  & spectra (ALEPH), multiplicities (PDG), and B hadron \\
  & fragmentation functions (ALEPH). \\
\hline
6 & second tune to LEP data by Nadine Fischer (September 2013). \\
  & Similar to the first one, but event shapes are weighted up \\
  & significantly, and multiplicities not included. \\
\hline
7 & (default) the Monash 2013 tune by Peter Skands at al.~\cite{MONASH-2013}, to \\
  & both $\eepm$ and $\mathrm{pp}/\mathrm{p\overline{p}}$ data. \\
\hline
\end{tabular}

\caption{Available $\eepm$ parton shower tune choices in Pythia8~\cite{PYTHIA8}.}

\label{tab:tunes}

\end{center}
\end{table}

We decouple the statistical fluctuations in the comparison between
Pythia tunes by studying the changes in the ratio between the measured
asymmetry at the reconstruction level and the bare asymmetry $\AFBzero$
calculated at the generator level on the same event sample: 

\begin{eqnarray}
{\cal R}_\text{QCD effects} \equiv \left.\frac{\AFB(\text{measured})}{\AFBzero(\text{generator level})}\right|_\text{same events} \; .
\end{eqnarray} 

$\AFBzero$ is the asymmetry corrected by QED effects, without the
inclusion of any QCD corrections, and its central value is unaffected
by acollinearity cuts within the targeted precision. Therefore, the
proposed ratio tracks in the most precise way distortions of the
asymmetry that are exclusively due to QCD effects. 

\begin{figure}[htbp]
\centering
\includegraphics[width=\linewidth]{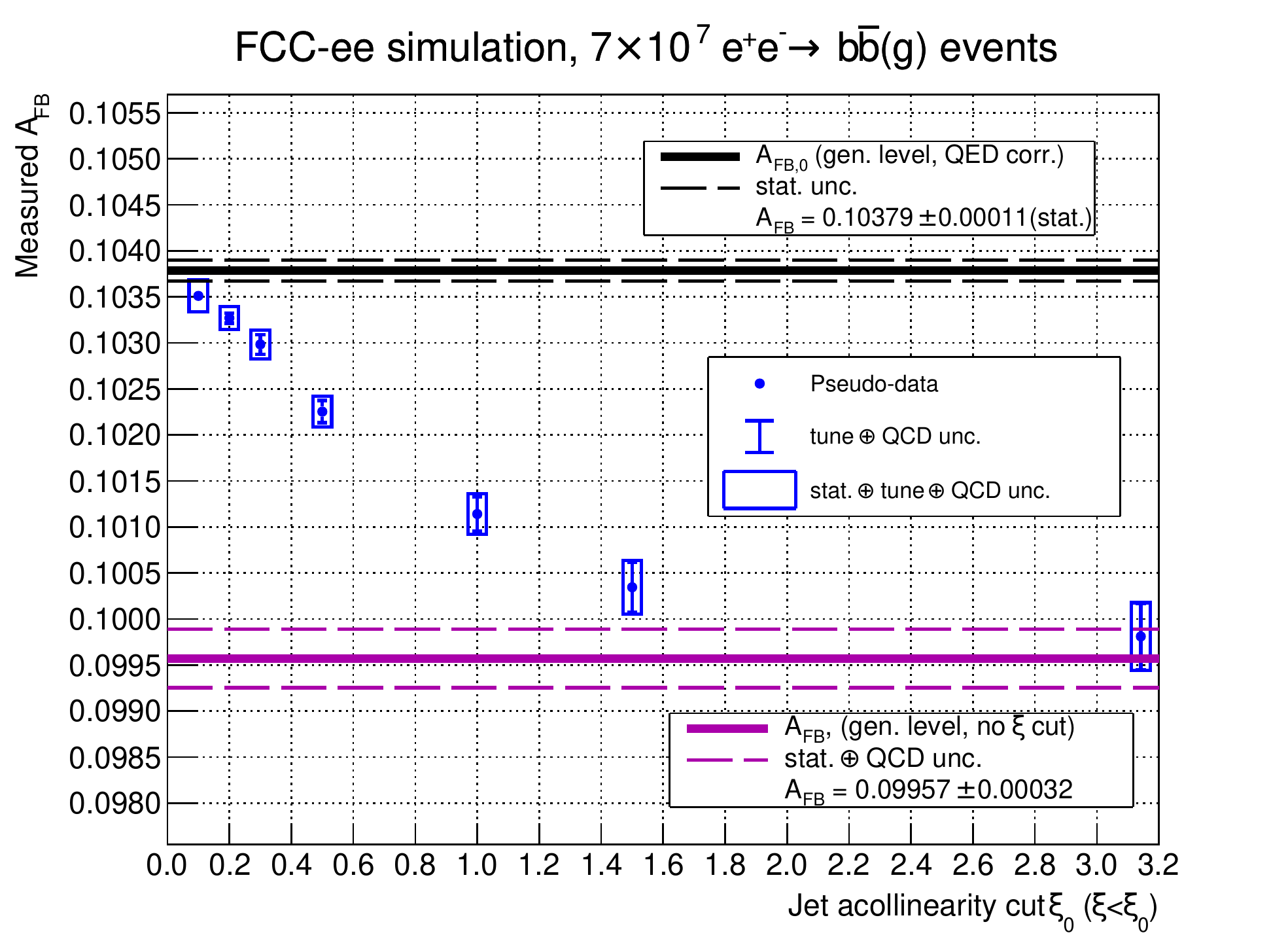}

\caption{Evolution of the expected measurement of the asymmetry in
$7\times 10^7 \eepm\to\bbbar(\mathrm{g})$ simulated events at
$\rts=91.2$~GeV as a function of the acollinearity cut $\xi<\xi_0$. The
systematic uncertainties in the pseudo-data points take into account
the envelope of the results obtained using different PYTHIA8 tunes,
which define the hadronization and timelike-showering properties in the
event generation. These results indicate that cuts in the range
$\xi_0=0.2-0.3$ should minimize the overall uncertainty. The
generator-level values of the asymmetries without and with QCD
corrections in the absence of acollinearity cuts are also shown for
reference.  \label{fig:afb-evolution}}

\end{figure}

With this new definition in hand we are now able to understand in more
depth the evolution of the asymmetry measurement as a function of the
$\xi_0$ cut applied at the reconstruction level. The central values of
the measurement are obtained by scaling the ratio ${\cal R}_\text{QCD
effects}$ at each point by the factor $\AFBzero = 0.10379 \pm
0.00011(\mathrm{stat.})$, determined using all the available
statistics.  Regarding uncertainties, three separate components are
considered at each $\xi_0$ point: 1) the statistical uncertainty
provided by the weighted average of the 7 tune results, basically
corresponding to $1/\sqrt{7}$ times the uncertainties reported in
Table~\ref{tab:afb-rec} for the Monash 2013 tune case; 2) the
(symmetrized) envelope of the central results obtained using different
tunes; 3) a theoretical uncertainty equivalent to a $10\%$ relative
uncertainty on the correction factor $C$, consistent with the
uncertainty assumed in LEP measurements~\cite{lephf-05}. Central values
and uncertainties are collected in Table~\ref{tab:afb-tune}, and
depicted in Figure~\ref{fig:afb-evolution}. Results are also corrected
for small biases due to the limited angular resolution effects, as
discussed in Section~\ref{sec:bias}. 

\begin{table}[htb]
\begin{center}

\begin{tabular}{|c|c|c|c|c|} \hline
$\xi_0$ cut & Measured $\AFB$ & $\Delta\AFB(\text{stat})$ & $\Delta\AFB(\text{tune})$ & $\Delta\AFB(\text{theo. QCD corr})$ \\
\hline\hline
No cut & $0.0998 \pm 0.0004$ & $0.00008$ & $0.00014$ & $0.00033$ \\
$1.50$ & $0.1003 \pm 0.0003$ & $0.00011$ & $0.00014$ & $0.00023$ \\
$1.00$ & $0.1011 \pm 0.0002$ & $0.00011$ & $0.00010$ & $0.00016$ \\
$0.50$ & $0.1023 \pm 0.0002$ & $0.00011$ & $0.00010$ & $0.00007$ \\
$0.30$ & $0.1030 \pm 0.0002$ & $0.00011$ & $0.00010$ & $0.00003$ \\
$0.20$ & $0.1033 \pm 0.0001$ & $0.00011$ & $0.00005$ & $0.00002$ \\
$0.10$ & $0.1035 \pm 0.0002$ & $0.00016$ & $0.00005$ & $0.00001$ \\
\hline
\end{tabular}

\caption{Central values and components of the uncertainty in the
measurement of the $\AFB$ asymmetry with $7\times 10^7$
$\eepm\to\bbbar(\mathrm{g})$ events at the Z pole, for different
$\xi<\xi_0$ cuts at the reconstructed level.}

\label{tab:afb-tune}

\end{center}
\end{table}

Changes in the central values of the asymmetry as a function of the
acollinearity cut can be largely explained by the different size of the
theoretically expected QCD corrections at each point. The uncertainty
on these corrections is larger ($\Delta\AFB(\text{QCD corr}) \approx
0.0003$) when no acollinearity cuts are applied. Pythia tune
uncertainties seem to have a marginal effect for $\xi_0<0.5$
($\Delta\AFB(\text{tune}) \lesssim 0.0001$) and are relatively stable
down to rather low values of the acollinearity cut. Statistical
uncertainties start to dominate for $\xi_0\lesssim 0.3$, but let us
remind that the statistical uncertainty will not be a limiting factor
at FCCee, where $\approx 10^{12}$ Z decays should be available. We
conclude that, for a real analysis of $\eepm\to\bbbar(\mathrm{g})$
events, a cut $\xi<0.2-0.3$ is already optimal, with associated QCD
systematics $\Delta\AFB(\text{tune+QCD corr}) \lesssim 0.0001$.

Figure~\ref{fig:afb-evolution} also shows the generator-level reference
values of the asymmetry with and without QCD corrections, calculated in
the absence of acollinearity cuts. Let us note again that we only
expect a qualitative agreement with the reference value with QCD
corrections.  For instance, there are hidden implicit cuts on
acollinearity at the selection level. We consider reconstructed jets
with a limited resolution parameter ($0.4$), but require at least two
tagged b-jets in the event.  Even without any explicit cut, the
double-tag requirement rejects events in which bottom and antibottom
quarks are in the same jet. Moreover, this region of phase space is
affected by gluon splitting contamination, something difficult to
reproduce in simulations. All of this encourages even further the use
of explicit acollinearity cuts and double flavor tagging in the
analysis before estimation of any QCD corrections.

\section{QCD uncertainties and b semileptonic decays}

One of the proposals to reduce QCD uncertainties was the one discussed
in Reference~\cite{ep-98}. The authors of the study analyzed the
evolution of the QCD corrections as a function of the momentum and the
transverse momentum of leptons (with respect to the associated jet
direction) in b decays. The corrections are quantified in terms of the
parameter $C_b = \frac{\as}{\pi}~C$ or, equivalently:

\begin{eqnarray}
C_b  & \equiv 1 - \frac{\AFB}{\AFBzero} \; .
\end{eqnarray}

\begin{figure}[htbp]
\centering
\includegraphics[width=\linewidth]{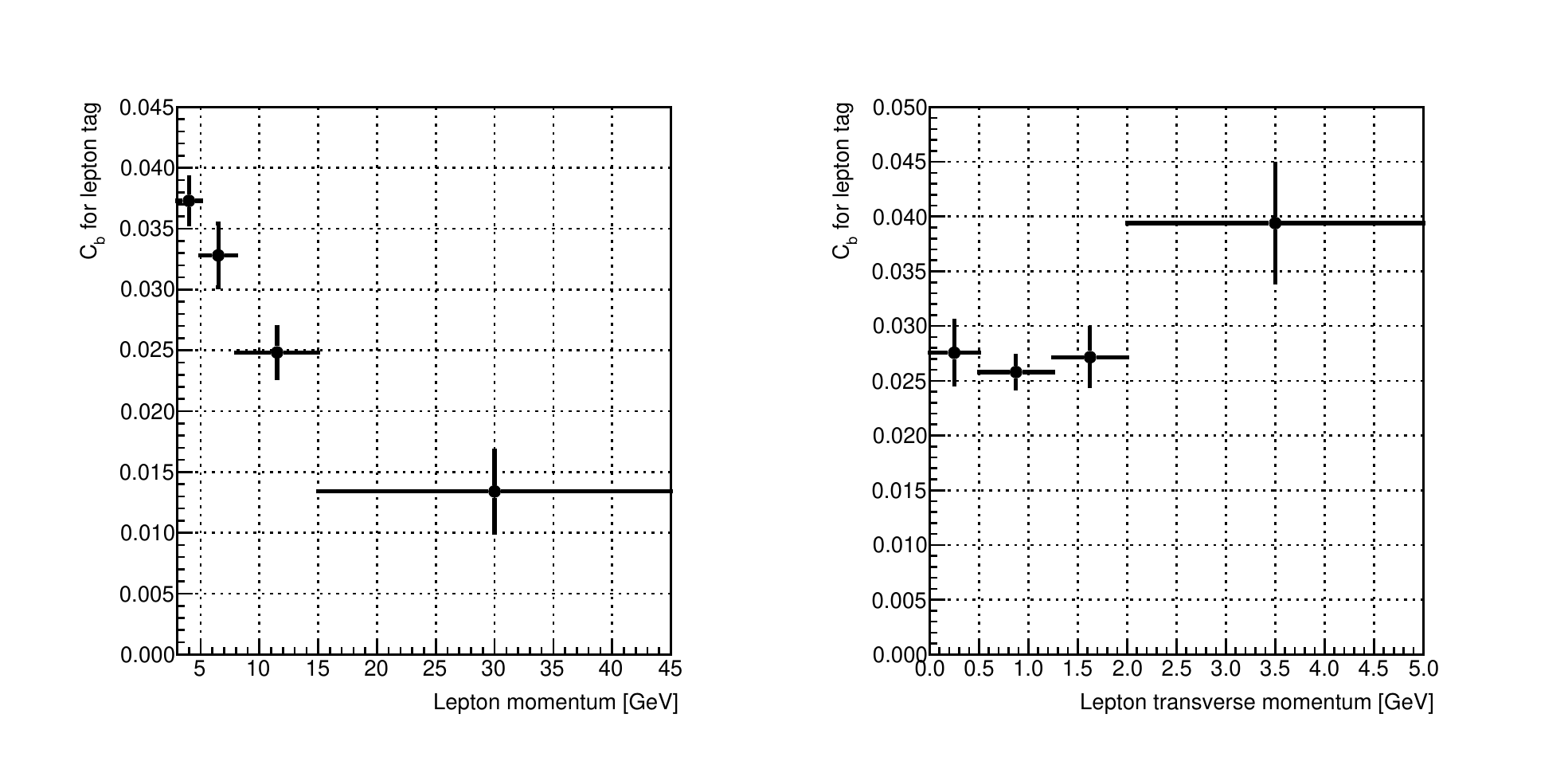}

\caption {QCD corrections factors $C_b$ in events with charged leptons
in the final state. Only leptons with a momentum larger than 3 GeV are
considered in the estimates. The results are presented as a function of
the lepton momentum and transverse momentum with respect to the closest
b-tagged jet direction. The central values of $C_b$ in each bin are
obtained as the average of the estimates of the 7 different Pythia tune
samples, while uncertainties are derived from the root-mean-square
deviation of these estimates.  \label{fig:cb}}

\end{figure}

We estimate $C_b$ using the $7\times 10^7$ simulated events analyzed in
Section~\ref{sec:tunes}. In order to be more consistent with the
analysis of Reference~\cite{ep-98}, we select events with a charged
lepton in the final state at generator level with momentum larger than
3 GeV. The results are shown if Figure~\ref{fig:cb}. The central values
of $C_b$ in each bin are obtained as the average of the estimates of
the 7 different Pythia tune samples, while uncertainties are derived
from the root-mean-square deviation of these estimates. Our findings
are qualitatively consistent with those reported in~\cite{ep-98},
except at large values the transverse momentum, probably due to the
presence of looser requirements in the angular association between
lepton and jet for the present study.

\begin{figure}[htbp]
\centering
\includegraphics[width=\linewidth]{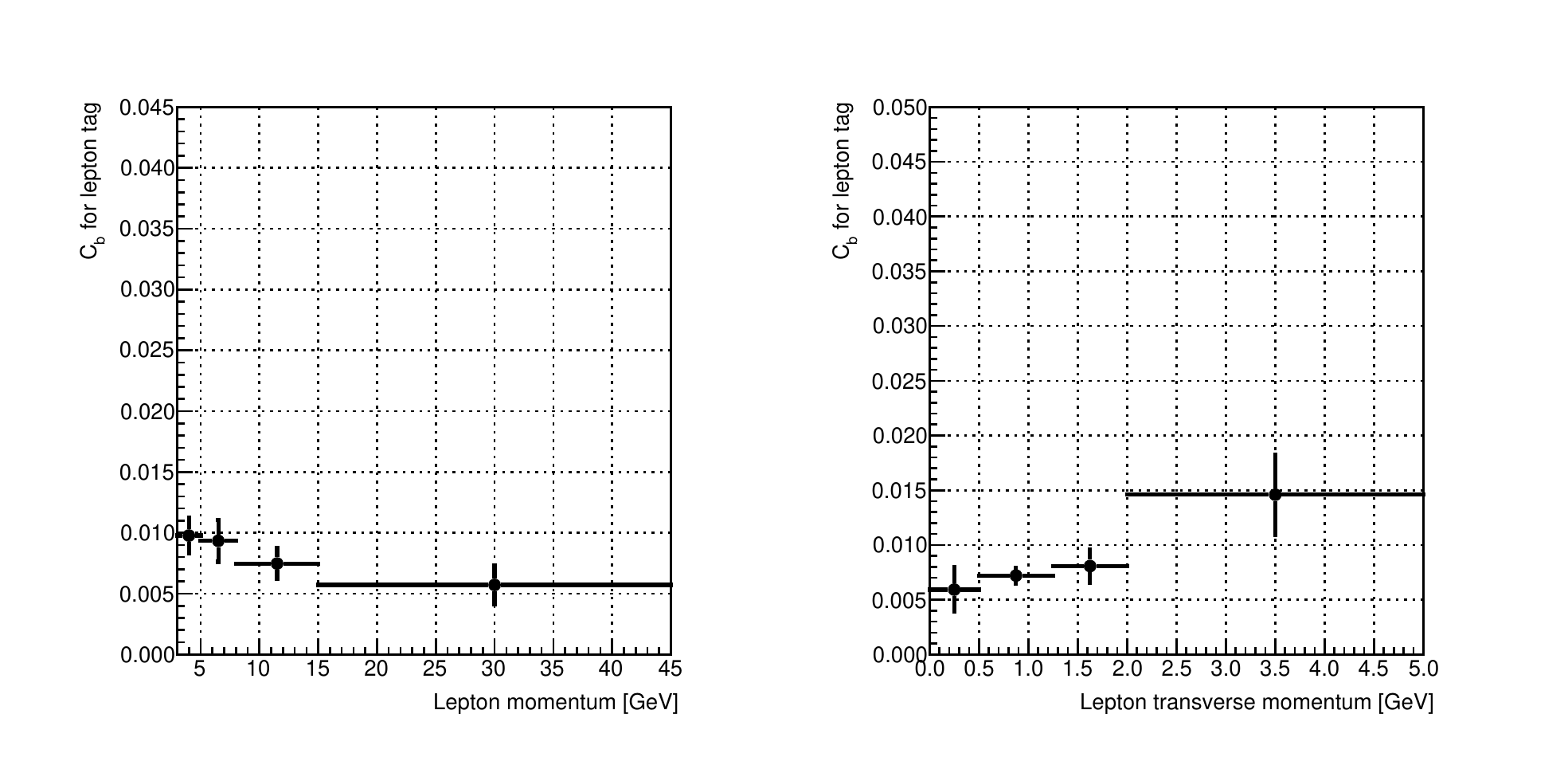}

\caption {QCD corrections factors $C_b$ in events with charged leptons
in the final state.  Only leptons with a momentum larger than 3 GeV are
considered in the estimates.  We require in addition the presence of
two reconstructed b-tagged jets with an acollinearity cut $\xi<0.3$.
The results are presented as a function of the lepton momentum and
transverse momentum with respect to the closest b-tagged jet direction.
The central values of $C_b$ in each bin are obtained as the average of
the estimates of the 7 different Pythia tune samples, while
uncertainties are derived from the root-mean-square deviation of these
estimates.  \label{fig:cb03}}

\end{figure}

The results are easily interpreted in terms of the acollinearity
discussion of the previous sections.  Tight cuts on the lepton momentum
reduce significantly the number of events with hard gluon radiation
close to that jet and, as a consequence, increase the number of
collinear events. But the reduction is less effective than the one
using explicit acollinearity cuts, simply due to the absence of
requirements on the opposite jet. This is visually shown in
Figure~\ref{fig:cb03}, using the same scale for comparison. Here we
require in addition the presence of two reconstructed b-tagged jets
with an acollinearity cut of $\xi<0.3$. The size of the QCD corrections
gets significantly reduced.  Also as expected, tight cuts on transverse
momentum lead to an increase in the size of QCD corrections: low values
are more consistent with a collinear configuration (low values of
$C_b$), while high values of transverse momentum favor configurations
with hard gluon emission away from the jet, as also observed by the
authors of the original study~\cite{ep-98}.

\section{Biases in the measurement of the asymmetry. Choice of the differential angular distribution. \label{sec:bias}}

Biases in the measurement of forward-backward asymmetries are
intrinsically small. Efficiency losses increase the statistical
uncertainty of the measurement, but do not bias it, as long as they are
independent of charge and a proper fit method is used. However, at the
level of the precision expected at future colliders, a real study with
data will have to carefully include all potential sources of bias.
Here we will focus in the particular case of $\AFBzerob$, but similar
considerations are applicable to the case of charm quarks. We will
distinguish four different types of bias: 1) charge confusion effects,
like B-mixing or experimental charge misidentification; 2) angular
distortions, and in particular the effect of a limited angular
resolution; 3) background contamination (mostly charm and light quarks)
; 4) wrong description of the polar angular distribution.

Charge confusion leads to a dilution of the asymmetry. The functional
form of the angular distribution does not change if the charge
confusion has no significant angular dependence:

\begin{eqnarray}
      \frac{d\rho}{d\cosb} & = & \frac{1}{1+\longPar}~\left[\frac{3}{8}~(1+\cosbSq)+\frac{3}{4}~\longPar\sinbSq \right]+\AFBprime~\cosb \; ,
\end{eqnarray}

$\AFBprime$ must be corrected a posteriori to obtain the true value of
the asymmetry.  Assuming a fraction $\chi$ of events with the wrong
charge: $\AFBprime = (1-2\chi)~\AFB$. In the more general case of a
symmetric dependence with the polar angle: $\chi\equiv \chi(|\cosb|)$,
a slightly different distribution must be used:

\begin{eqnarray}
      \frac{d\rho}{d\cosb} & = & \frac{1}{1+\longPar}~\left[\frac{3}{8}~(1+\cosbSq)+\frac{3}{4}~\longPar\sinbSq \right]+\AFB~[1-2\chi(|\cosb|)]~\cosb \; . \label{eq:full-angular-distribution}
\end{eqnarray}

The function $\chi(|\cosb|)$ can be extracted either from simulations or
from dedicated studies in control data samples. It can also be directly
measured from a global data fit with multiple charge and flavor tags.
The remaining biases are expected to be small. At LEP, B-mixing was
already an almost negligible source of systematic uncertainty in the
combined measurement ($\Delta\AFBzerob \approx 0.0001$).
Detector-related charge confusion effects can be estimated in control
samples with double tagging of the charge, and is expected a priori to
be small in several cases of interest, like measurements that use
exclusive decays or inclusive semi-leptonic b decays~\cite{lephf-05}.

The limited angular resolution in the polar angle, $\sigmaTh$, is
intrinsically a source of a bias in the measurement. Any value of
$\sigmaTh>0$ flattens the original angular distribution and results
into a lower absolute value of the fitted asymmetry. To illustrate the
effect of a typical jet resolution we generate a toy sample of $10^9$
events with the resolution parameters extracted from
Figure~\ref{fig:angular-reso}-right of Section~\ref{sec:afb-rec}, for
an acollinearity cut of $\xi<0.3$. We perform a likelihood fit to the
usual angular distribution (which assumes $\sigmaTh=0$) in the fiducial
volume $|\cosb|<0.97$, consistent with the tracker acceptance of the
IDEA detector. The result is a relatively small shift of
$\Delta\AFB=-0.000195\pm 0.000014$, which is used to correct the
results obtained with likelihood fits in previous sections. Let us note
that the angular resolution can still be improved by combining the jet
angular information with the measurement of the direction of the
secondary vertex, which is much closer to the original b-quark
direction.  Nevertheless, we conclude that angular resolution effects
should be studied and carefully corrected in a real analysis.

For the b-quark asymmetry, and at the level of the aimed precision, the
contamination from charm and light quarks can be estimated in situ
using control samples, or even extracted directly from data. A very
precise extraction of the background will likely require an extension
to double flavor-tag methods, the addition of flavor-sensitive
variables, and the formulation of complicated probability density
functions to appropriately describe all the flavor components. In any
case, in LEP combinations the effect of uncertainties in the background
composition was already minor: $\Delta\AFB\approx
0.0001$~\cite{lephf-05}. Improved heavy-quark tagging capabilities and
the larger collected statistics at future colliders should provide an
even better control of these contributions.  Also, with $\approx
10^{12}$ Z decays, one could envisage measurements using only a few
exclusive channels.  For instance, $\approx 10^8$ $\mathrm{B}^+$ decays
will be available~\cite{Fabrizio} for a measurement minimally affected
by B-mixing, charm/light background or charge confusion.

The last source of bias that we consider is the choice of the angular
distribution used to fit the asymmetry. A measurement with huge
statistics will certainly demand careful tests of its validity. For
instance, a measurable longitudinal component due to new physics can
not be excluded a priori, given the level of precision. An example
would be the exchange of a scalar particle in the s-channel with
sizable couplings to electrons and heavy quarks or, in a more generic
way, the presence of an effective interaction of the
$(\overline{\mathrm{e}}_L \mathrm{e}_R) (\overline{\mathrm{b}}_L
\mathrm{b}_R)$ type. In this context, one could even envisage a fit
with two parameters, $\longPar$ and $\AFB$, using
Equation~\ref{eq:full-angular-distribution} above.

Finally, we should point out that the usual description in terms of Z
decay angular components is not strictly valid when higher-order
electroweak corrections are taken into account. This is simply a
manifestation of quantum interference effects between Z production and
Z decay terms in the $\eeQQg$ process.  Even if these interference
effects are expected to be negligible at the Z pole, they become larger
as we move away from this energy point. As an example, initial-final
state QED interference effects, critical for a precise measurement of
the asymmetry in the $\eepm\to\mu^+\mu^-(\gamma)$ process, have been
investigated in Reference~\cite{Jadach:2018lwm}. The conclusion is that
effects as large as $1\%$ can be present at energies a few GeV above
the Z mass. In practice, accounting for these effects should not be
extremely problematic. One can assume a more general polar angle
distribution, with symmetric and antisymmetric density terms
$\mathcal{S}(\cosbSq)$ and $\mathcal{A}(\cosb)$, and properly
normalized such that: 

\begin{eqnarray}
      \frac{d\rho}{d\cosb} & = & \mathcal{S}(\cosbSq) + \AFB~\mathcal{A}(\cosb) \;.
\end{eqnarray}

The exact values of the density terms as a function of $\cosb$ can be
even obtained from analytical theory calculations. The solution of the
unbinned likelihood fit to this distribution is:

\begin{eqnarray}
      v_i \equiv \frac{\mathcal{A}(\cosb{_{,i}})}{\mathcal{S}(\cosbSq{_{,i}})} & \Rightarrow & 
      \sum_{i=1}^{N_{events}} \left( \frac{v_i}{1 + \AFB~v_i}\right) = 0 \; ,
\end{eqnarray}

\noindent which is still unbiased with respect to acceptance variations
that are charge symmetric. The consistency of the functional form can
be confirmed a posteriori through a study of the dependence with $|v|$
of the following function:

\begin{eqnarray}
f(|v|) & = & {\displaystyle \frac{\left[dN(|v|)-dN(-|v|)\right]} {\left[dN(|v|)+dN(-|v|)\right]} } \; ,
\end{eqnarray}

\noindent where $dN(x)$ is the number of selected events in the
$(x-\frac{dx}{2},x+\frac{dx}{2})$ interval. If the functional form is
appropriate then one expects a good fit of the histogram of $f(|v|)$
versus $|v|$ to the linear relation: $f(|v|) = \AFB~|v|$. Actually, an
alternative value of $\AFB$ can be extracted from this fit. Note
however that the resulting statistical uncertainty will be by
construction larger than the optimal one, which is given by the
unbinned maximum likelihood fit described before.

\section{Outlook}

Using existing theoretical calculations as a starting point we have
derived simple expressions for the QCD corrections to the measurement
of the forward-backward asymmetry of the $\eeQQg$ process at
$\rts=\MZ$. We show that these QCD corrections and uncertainties can be
reduced by almost one order of magnitude using simple acollinearity
cuts between the heavy quarks. The inclusion of hadronization and
final-state parton shower effects does not modify this picture. These
findings are also enough to explain the behavior of other observables
previously suggested for a partial mitigation of QCD effects. The
proposed strategy is a first step towards a measurement of the
heavy-quark forward-backward asymmetry with systematic uncertainties of
order $\Delta\AFBzeroQ \approx 10^{-4}$.  Focusing on the case of b
quarks, it is found that other possible sources of bias in the
measurement of the asymmetry could be controlled at that level of
precision using current theoretical knowledge, adequate fitting
techniques and an appropriate choice of data control samples.

\section*{Acknowledgments}

We would like to thank Patrick Janot and Emmanuel Perez for useful
discussions, suggestions and a detailed reading of the manuscript.

\newpage
\bibliographystyle{myutphys}
\bibliography{AFBqcd-hep}{}

\end{document}